\documentclass[aps,prl,twocolumn,groupedaddress]{revtex4}

\usepackage{graphicx}
\usepackage{amsmath}
\usepackage{tabu}
\usepackage{booktabs}

\begin{document}

\title{Structural and magnetic phase transitions in EuTi$_{1-x}$Nb$_x$O$_3$}

\author{Ling Li$^{1}$, James Morris$^{1,2}$, Michael Koehler$^{1}$, Zhiling Dun$^{3}$, Haidong Zhou$^{3}$, Jiaqiang Yan$^{1,2}$, David Mandrus$^{1,2}$, Veerle Keppens$^{1}$}

\affiliation{$^{1}$Department of Materials Science and Engineering, The University of Tennessee, Knoxville, TN 37996, USA \\$^{2}$ Materials Science and Technology Division, Oak Ridge National Laboratory, Oak Ridge, TN 37831, USA\\$^{3}$Department of Physics and Astronomy, The University of Tennessee, Knoxville, TN 37996, USA}

\date{\today}

\begin{abstract}
We investigate the structural and magnetic phase transitions in EuTi$_{1-x}$Nb$_{x}$O$_{3}$ (0 $\le$ $x$ $\le$ 0.3) with synchrotron powder X-ray diffraction (XRD), resonant ultrasound spectroscopy (RUS), and magnetization measurements. 
Upon Nb-doping, the $Pm\bar{3}m$ $\leftrightarrow$ $I4/mcm$ structural transition shifts to higher temperatures and the room temperature lattice parameter increases while the magnitude of the octahedral tilting decreases. In addition, Nb substitution for Ti destabilizes the antiferromagnetic ground state of the parent compound and long range ferromagnetic order is observed in the samples with $x \ge$ 0.1. The structural transition in pure and doped compounds is marked by a step-like softening of the elastic moduli in a narrow temperature interval near $T_S$, which resembles that of SrTiO$_3$ and can be adequately modeled using the Landau free energy model employing the same coupling between strain and octahedral tilting order parameter as previously used to model SrTiO$_3$. 

\end{abstract}


\maketitle

\section{INTRODUCTION}
The tilting of the oxygen octahedra in cubic perovskites is known to induce structural phase transitions, which are often associated with the emergence of intriguing physical phenomena \cite{Imada1998}. 
SrTiO$_3$ (STO) is one of the most extensively studied perovskite oxides for its structural phase transition (SPT) at $T_S$ $\approx$ 105 K \cite{Fleury1968,STOinelastic}. 
This SPT has been recognized as the paradigm of a displacive transition, nevertheless, proven to carry some order/disorder element \cite{STOdisorder,Bussmann-Holder2007}. 
In the past decade, the discovery of the magnetoelectric coupling in iso-structural EuTiO$_3$ (ETO) has generated substantial interest in this compound \cite{Katsufuji2001}. This coupling manifests itself in a significant decrease of the dielectric constant at the G-type antiferromagnetic ordering temperature $T_N$ $\approx$ 5.6 K \cite{Katsufuji2001}. 
In analogy to STO, ETO also undergoes an antiferrodistortive SPT from $Pm\bar{3}m$ to $I4/mcm$ ($a^0a^0c^-$ in Glazer notation) driven by R-point phonon softening \cite{Inelastic_X-ray_ETO}. 
While the SPT at 105 K in STO has been extensively studied for decades, the transition in ETO was observed only recently \cite{Bussmann-Holder2011} but has since attracted much attention in the field and has been the focus of numerous experimental \cite{Taylor2012,Allieta2012,Goian2012,Zurab2012,Inelastic_X-ray_ETO,satellite,NoSatellitePeak,Bessas2013} and theoretical \cite{Rushchanskii2012,Yang2012,PhysRevB.88.094103} studies.

Previous synchrotron powder X-ray diffraction (XRD) indicates a $T_S \approx$ 235 K for ETO \cite{Allieta2012}, which is significantly lower than $T_{A}$ $\approx$ 282 K suggested by heat capacity measurement \cite{Bussmann-Holder2011}. 
This discrepancy draws attention to the possible local tetragonal distortions present above 235 K, as evidenced by pair distribution function analysis \cite{Allieta2012}. 
Kim \textit{et al.} \cite{satellite} identified incommensurate satellite peaks below 285 K in single crystal synchrotron XRD, and proposed a dynamically modulated equilibrium state between 285 K and 160 K, incorporating antiferrodistortive instabilities of TiO$_6$ octahedra accompanied by antiferroelectric displacements of Ti atoms. 
Conversely, a simple and un-modulated structure is claimed for ETO, as evidenced by the commensurate superlattice peaks at $T$ = 250 K in single crystal XRD \cite{NoSatellitePeak}, and the R-point acoustic phonon softening that is similar to the soft modes in STO \cite{Inelastic_X-ray_ETO}.    
ETO samples synthesized under different conditions were studied \cite{Goian2012}. 
While one of the ceramic samples shows a simple cubic-to-tetragonal long range transformation near 300 K, the other ceramic samples as well as the single crystal present an incommensurate structure at low temperature. 
These discrepancies were considered to arise from variations in sample preparation conditions, which lead to mixed valence on the Eu-site and/or oxygen non-stoichiometry and therefore could potentially alter the physical properties. 
This was further confirmed by Kennedy \text{et al.} \cite{Kennedy2014}, who show that the electrical resistivity of EuTiO$_{3-\delta}$ is significantly reduced with oxygen vacancies due to the change of Ti$^{4+}$ to Ti$^{3+}$ when oxygen vacancies and/or trivalent Eu are present. 
Very recently, the elastic response near 1 MHz of an irregularly shaped ETO single crystal was examined using resonant ultrasound spectroscopy (RUS) \cite{Spalek2014}. A pronounced step-like shear softening near $T_S$ $\approx$ 284 K confirms that the SPT in STO and ETO is very similar in nature but takes place at very different temperatures.

Despite the recent studies of the nature and $T_{S}$ of the SPT in ETO, the dramatic difference in $T_S$ between STO and ETO is not well understood (the Goldschmidt tolerance factor $t$ predicts a similar $T_S$ for these two compounds) and may signal the importance of the spin-phonon coupling in ETO \cite{Lee2010}.
First-principles calculations suggested that the hybridization between Eu-$f$ and Ti-$d$ orbitals enhances the tilting of the octahedra and therefore $T_{S}$ \cite{PhysRevB.88.094103}. 
To examine how the $T_{S}$ of the ETO system is influenced by the hybridization between Eu-$f$, (Ti, Nb)-$d$, and O-$p$ orbitals, we have initiated the study of EuTi$_{1-x}$Nb$_{x}$O$_{3}$. Nb doping introduces itinerant electrons into this system, which mediate the magnetic interactions between Eu 4$f$ spins and result in ferromagnetism \cite{Li2014}. 
In the present paper, we extend the RUS study by Spalek \textit{et al.} \cite{Spalek2014} and report the full elastic tensor of single crystals of ETO as well as mixed crystals of EuTi$_{1-x}$Nb$_x$O$_3$. The dramatic change in the elastic constants allows for an accurate determination of $T_S$, and the potential impact of the orbital hybridization on $T_S$ can be addressed. The amplitude of octahedral rotations at 100 K is determined from synchrotron X-ray powder diffraction data, and the role of Nb doping on the structural and magnetic instabilities will be discussed.

\section{Methods}
EuTi$_{1-x}$Nb$_x$O$_3$ polycrystals (0 $\le x \le$ 0.3) and single crystals ($x$ = 0, 0.05, 0.1 and 0.2) have been synthesized. 
The polycrystalline samples were prepared using a conventional solid-state reaction method as described in Ref. \cite{Li2014}. 
The single crystal growth was carried out in a double-ellipsoidal-mirror image furnace (NEC) in 90:10 Ar:H$_2$ atmosphere. 
The pulling speed was 10 mm/h with feed and seed rods rotating in opposite directions at 25 rpm. Good crystal quality was confirmed by clear, round spots in the Laue back scattering pattern. The single crystals were oriented with $Pm\bar{3}m$ symmetry using the software OrientExpress and cut to rectangular parallelepipeds with all faces perpendicular to the crystallographic $\langle100\rangle$ axes for RUS measurements. The edge dimensions of the EuTi$_{1-x}$Nb$_x$O$_3$ ($x$ = 0, 0.05, 0.1) single crystals for RUS measurements range between 0.5 mm and 2.5 mm. 
RUS is a technique developed by Migliori \textit{et al.} \cite{Migliori19931} for determining the complete elastic tensor of a small single crystal by measuring its free-body resonances. It allows determination of the full elastic tensor without needing to change transducers or remounting the sample.

Room and low temperature X-ray diffraction (XRD) was first performed with a laboratory HUBER Imaging Plate Guinier Camera 670 with Ge monochromatized $Cu K_{\alpha 1}$ radiation ($\lambda = 1.54059 {\AA}$), confirming all samples to be single phase. 
To better resolve the tetragonal distortion, synchrotron powder XRD was performed through a mail-in program at the 11-BM beamline of the Advanced Photon Source at Argonne National Laboratory.    
Single crystals of EuTi$_{1-x}$Nb$_x$O$_3$ ($x$ = 0, 0.05, 0.1 and 0.2) were ground to fine powders and loaded in a capillary tube with a diameter of 0.8 mm. 
The experiments were carried out with a fixed wavelength ($E$ = 30 $keV$, $\lambda = 0.41396(1) {\AA}$) at $T$ = 295 K and 100 K. 
Rietveld refinements were obtained using the program GSAS \cite{EXPGUI}. 

The magnetization measurements were performed using a Quantum Design Magnetic Property Measurement System in the temperature interval 2 K to 300 K after cooling in either zero field (ZFC) or in a measuring field (FC).
Thermogravimetric analysis (TGA) was employed to determine the oxygen content of ETO from the weight gain due to oxidation of Eu$^{2+}$ to Eu$^{3+}$. The ETO crystals were found to be oxygen stoichiometric within experimental error.

\section{results}
\subsection{A. X-ray diffraction}
Fig. 1(a) and (b) show the Rietveld refinements from synchrotron XRD for the parent ETO at 295 K and 100 K, respectively. 
While the pattern at 295 K was refined with the $Pm\bar{3}m$ space group with lattice parameter $a$ = 3.9047(3) ${\AA}$, the pattern at 100 K was refined using $I4/mcm$ symmetry, yielding lattice parameters $a_{Tetra} = \sqrt{2}a_{Cub} = 5.51168(3) {\AA}$ and $c_{Tetra} = 2c_{Cub} = 7.81613(5) {\AA}$, in good agreement with previously reported values \cite{Allieta2012}.
Furthermore, the 295 K synchrotron XRD patterns of all samples show no well-defined peak splitting and could be adequately fitted with a cubic $Pm\bar{3}m$ model, whereas the 100 K patterns show well-resolved peak splitting and were fitted with a tetragonal $I4/mcm$ model.
The room temperature lattice parameter extracted from Rietveld refinements obeys Vegard's law, and increases linearly with increasing $x$ in EuTi$_{1-x}$Nb$_{x}$O$_{3}$. 
The symmetry lowering at 100 K was evidenced by the (200) peak splitting highlighted in the insets to Fig. 1. 

\begin{figure}[h]
\centering
\includegraphics[trim = 0cm 0cm 0cm 0cm, clip, width=0.5\textwidth]{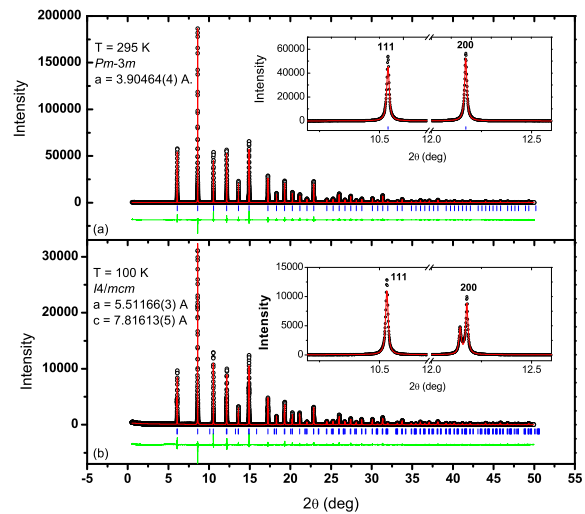}
\caption{Rietveld refinements of the synchrotron X-ray powder patterns of EuTiO$_{3}$ collected at (a) 295 K and (b) 100 K, respectively. The insets show the (111) and (200) reflections.}
\label{FIG.1}
\end{figure}

\begin{figure*}[htbp]
\centering
\includegraphics[trim = 0cm 9cm 0cm 0cm, clip, width=1\textwidth]{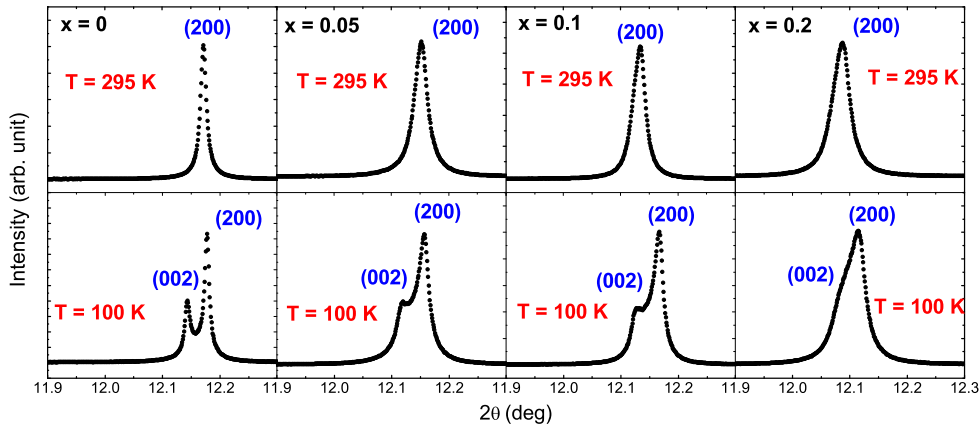}
\caption{(200) reflection at 295 K (top) and 100 K (bottom) for EuTi$_{1-x}$Nb$_x$O$_3$ ($x$ = 0, 0.05, 0.1 and 0.2).}
\label{FIG.2}
\end{figure*}

\begin{figure}[!h]
\centering
\includegraphics[trim = 0mm 0mm 0mm 0mm, clip, width=0.4\textwidth]{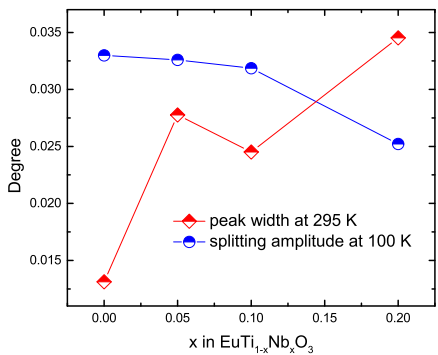}
\caption{The (200) peak width at 295 K and splitting amplitude at 100 K for EuTi$_{1-x}$Nb$_x$O$_3$ ($x$ = 0, 0.05, 0.1, 0.2).}
\label{FIG.3}
\end{figure}

\begin{figure}[!h]
\centering
\includegraphics[trim = 0mm 2cm 0mm 0mm, clip, width=0.5\textwidth]{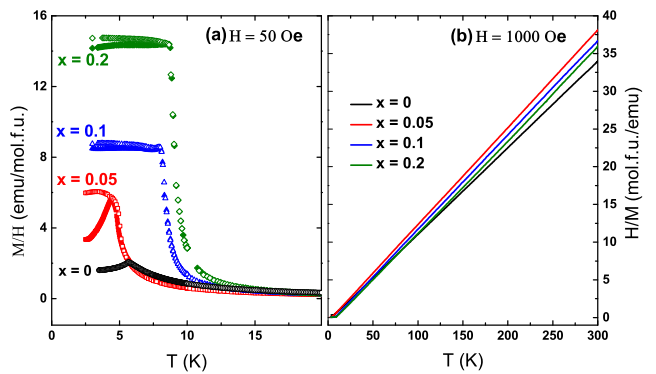}
\caption{Temperature dependence of magnetization for single crystals of EuTi$_{1-x}$Nb$_x$O$_3$ ($x$ = 0, 0.05, 0.1, 0.2). Note that the full and empty symbols in (a) denote ZFC and FC data, respectively.}
\label{FIG.4}
\end{figure}

As shown in Fig. 2, no readily apparent peak splitting due to the structural transition was observed at room temperature. 
However, we noticed significant peak broadening with increasing Nb content. 
This is demonstrated more clearly in Fig. 3 in which we plotted the peak width of the (200) reflection at 295 K and splitting amplitude of (200) at 100 K obtained by a Lorentzian fit. 
Nb doping induces disorder in the lattice, which can naturally explain the peak broadening. 
However, we believe that the peak splitting due to the structural transition also contributes to the peak broadening at 295 K since the SPT takes place above room temperature as revealed by RUS measurements. 
Intriguingly, the splitting amplitude at 100 K, which is proportional to the magnitude of the lattice distortion $c/a$ ratio, is suppressed with increasing Nb doping, suggesting that Nb doping weakens the octahedral tilting in the ETO system.

\subsection{B. Magnetization}
Fig. 4 shows the temperature dependence of magnetization for EuTi$_{1-x}$Nb$_x$O$_3$ single crystals, plotted as (a) M/H under 50 Oe, and (b) H/M under 1000 Oe. 
The ZFC and FC curves of the parent ETO overlap, and reveal a $T_N$ $\approx$ 5.7 K. The data under 1000 Oe between 200 K and 300 K could be well fitted by the Curie-Weiss law, yielding a positive Weiss temperature of 3.48(1) K and an effective moment of approximately 8.35(1) $\mu_B$, which agree well with literature values \cite{Katsufuji2001}. 
A dramatic divergence between ZFC and FC curves is present for $x$ = 0.05 sample, which may signal a spin glass behavior arising from the competition between AFM and FM interactions present in this system. 
The magnetic ground state of samples with $x \ge 0.1$ is switched from AFM to FM, as previously reported for polycrystalline samples \cite{Li2014}. 
It is noteworthy that Nb doping induces metallic behavior \cite{Li2014}.  
The induced ferromagnetism most likely results from the ferromagnetic interaction between localized Eu 4$f$ spins, mediated by itinerant electrons introduced by chemical doping \cite{Katsufuji1999,Li2014}.

\subsection{C. Elastic moduli}
Cubic systems exhibit three independent elastic constants $C_{11}$, $C_{12}$ and $C_{44}$. 
Note that $C_{11}$ and $C_{44}$ govern respectively longitudinal and transverse waves propagating along [100]. In the [110] direction, longitudinal waves are governed by $C_{L[110]} = \frac{1}{2}(C_{11}+C_{12}+2C_{44})$, while one transverse wave is governed by $C_{44}$, the other by $C_{T[110]} = \frac{1}{2}(C_{11}-C_{12}) = C'$. 
In Fig. 5 we plot the temperature dependence of $C_{44}$. 
Below $T_S$, the ultrasonic absorption of the sample is so large that not enough resonances can be observed to allow an accurate determination of all three elastic moduli. 
However, the high temperature fit indicates that the lowest resonant frequency depends almost exclusively on $C_{44}$. 
This frequency is visible throughout the entire transition and below, allowing determination of the shear modulus over the entire temperature range. 
The striking feature in Fig. 5 is the step-like elastic softening of $C_{44}$ at $T_S$ $\approx$ 288 K, which is reminiscent of the 105 K elastic anomaly observed in the elastic moduli of STO \cite{ElasticConstantsSTO}. 
The softening at $T_S$ is accompanied by a dramatic rise in the internal friction $Q^{-1}$ as a consequence of domain wall motion driven by elastic waves, which causes a significant broadening and deterioration of the resonances, preventing determination of the full tensor, as discussed above.

\begin{table*}[ht]
\centering
\caption{$T_S$, density, and elastic moduli for single crystals of STO, ETO and ETO-Nb.}
\label{my-label}
\begin{tabular}{cccccccc}
\hline
    Sample  & $T_S$ (K) & Theoretical Density $\rho$ (g/cm$^3$) & Density $\rho$ (g/cm$^3$) & $C_{11}$ (GPa) & $C_{44}$ (GPa) & $C'$ (GPa) & error $\%$  \\ \hline
STO (295 K)      &105 & 5.110                & 5.015           & 311.58    & 120.83    & 107.2    & 0.3507  \\
ETO (295 K) &288   & 6.914                 & 6.744           & 307.24    & 116.3     & 100.06   & 0.2323  \\
ETO (370 K)       &  288               &  6.914     &  6.744       & 309.76    & 116.95    & 102.57   & 0.2833  \\
ETO-5$\%$Nb (370 K) & 322  & 6.939                    & 6.809           & 316.07    & 112.56    & 104.365  & 0.1467  \\
ETO-10$\%$Nb (370 K) & 353 &     6.970                & 6.758           & 306.29    & 109.23    & 100.315  & 0.3138  \\ \hline
\end{tabular}
\end{table*}

\begin{figure}[ht]
\centering
\includegraphics[trim = 0cm 0mm 0mm 0cm, clip,width=0.5\textwidth]{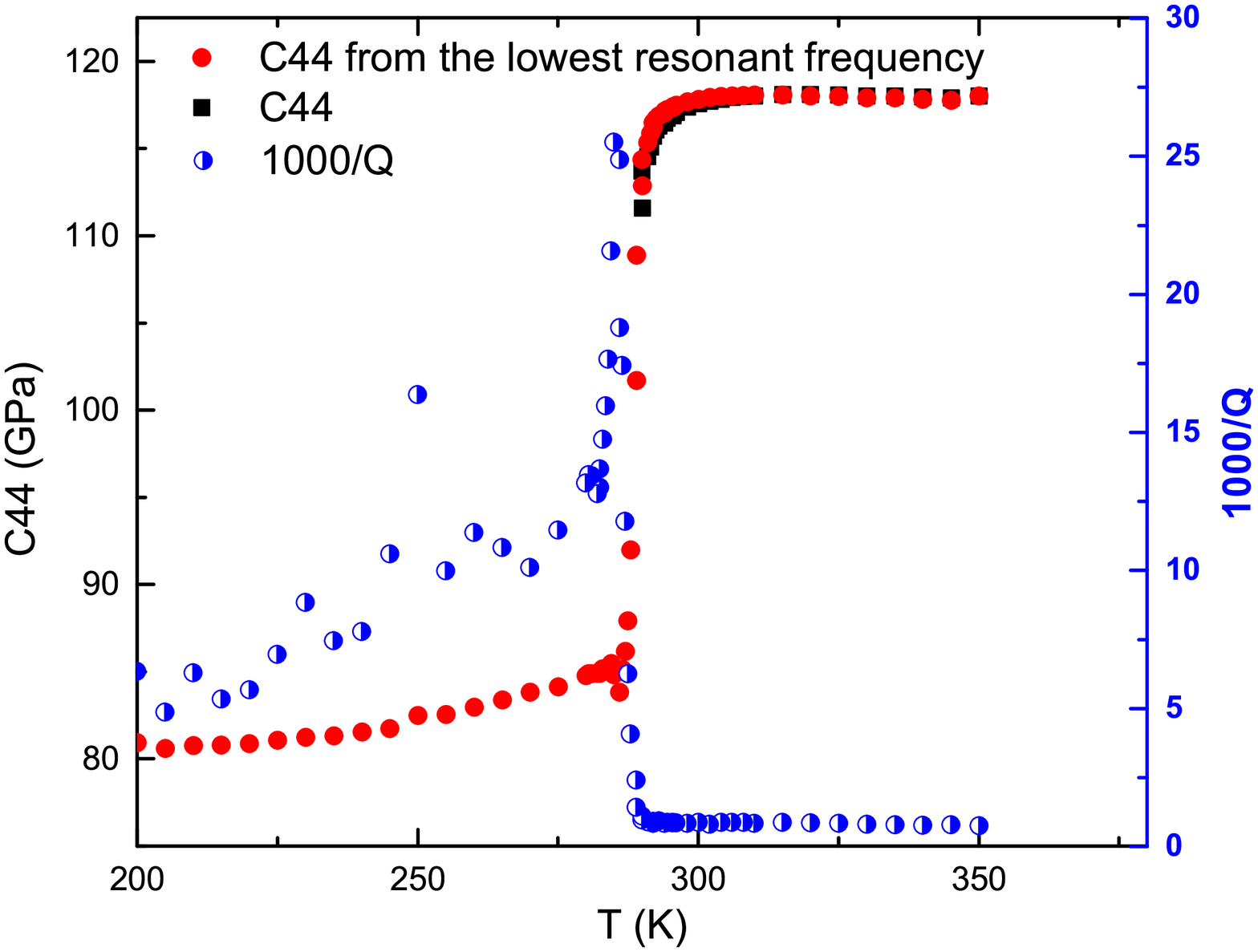}
\caption{Temperature dependence of the shear modulus $C_{44}$ and the inverse of the quality factor Q, which is defined as the center frequency of the resonance divided by full width at half maximum, of a oriented ETO single crystal. Note that below the structural phase transition temperature, $C_{44}$ is determined directly from the first resonance peak (red circles).}
\label{FIG.5}
\end{figure}

Fig. 6  presents the temperature dependence of $C_{11}$, $C_{44}$ and $C'$ for single crystals of EuTi$_{1-x}$Nb$_x$O$_3$ ($x$ = 0, 0.05 and 0.1) both in the absence of field and under applied magnetic fields up to 3 T. The main features are as follows:

1) The substitution of Nb$^{4+}$ for Ti$^{4+}$ (up to $x = 0.1$) does not change the qualitative behavior of the elastic response near the SPT, and the step-like elastic softening as the temperature approaches $T_S$ remains the most striking feature.

2) The transition temperature $T_S$ is shifted to higher temperatures with increasing Nb doping, $T_S$ = 288 K, 322 K and 353 K for $x$ = 0, 0.05 and 0.1, respectively.  

3) The elastic response of EuTi$_{0.95}$Nb$_{0.05}$O$_3$ does not show an observable field dependence up to 3 T. 

Table I shows the $T_S$, density $\rho$, and the values of the elastic moduli for various samples. The data for single crystal of STO is obtained from the present RUS study. It can be clearly seen that the absolute values of the moduli for STO and ETO are very close.

\begin{figure}[!h]
\centering
\includegraphics[trim = 2cm 3cm 8cm 0cm, clip,width=0.6\textwidth]{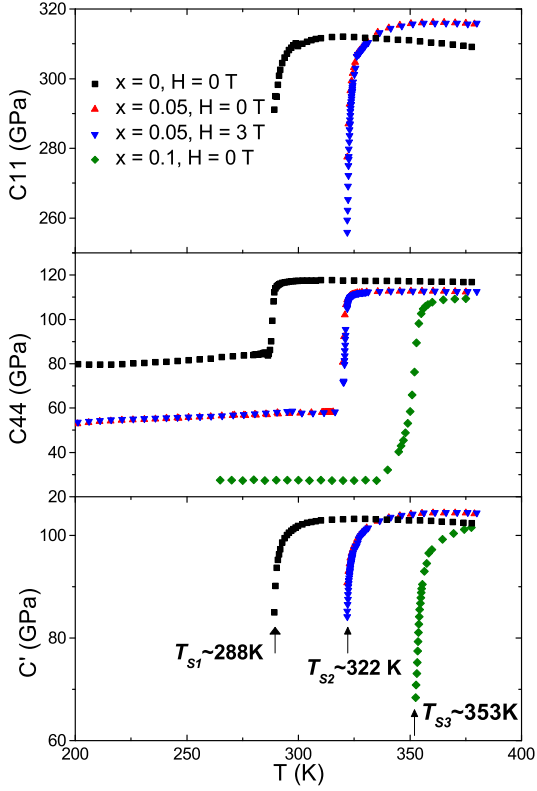}
\caption{Temperature dependence of elastic constants for single crystals of EuTi$_{1-x}$Nb$_x$O$_3$ ($x$=0, 0.05, and 0.1) in different magnetic fields. Measurements were performed during cooling.}
\label{Figure 6}
\end{figure}

\begin{figure}[!h]
\centering
\includegraphics[trim = 0mm 0cm 0cm 0mm, clip, width=0.5\textwidth]{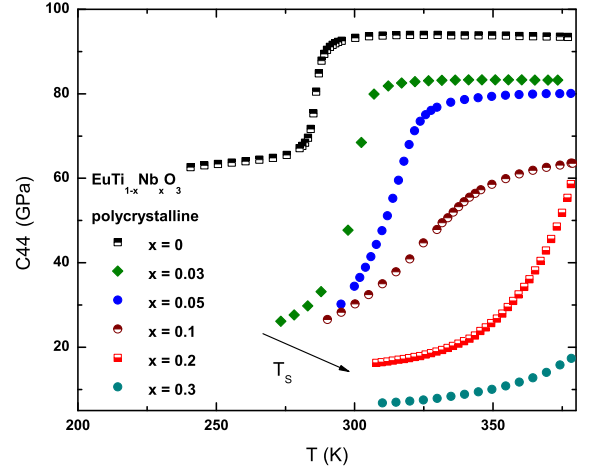}
\caption{Temperature dependence of the shear modulus $C_{44}$ for polycrystalline EuTi$_{1-x}$Nb$_x$O$_3$ ($x$=0, 0.03, 0.05, 0.1, 0.2, and 0.3). Data are shifted for clarity. Measurements were performed during cooling.}
\label{FIG.7}
\end{figure}

Fig. 7 shows the temperature dependence of $C_{44}$ for polycrystalline samples of EuTi$_{1-x}$Nb$_x$O$_3$ ($x$ = 0, 0.03, 0.05, 0.1, 0.2 and 0.3). 
Because the porosity of the ceramic sample changes the density, and therefore the absolute value of the elastic moduli, we are only interested in the qualitative behavior, and to improve clarity the curves are offset along the $y$ axis. 
The structural phase transition in EuTi$_{1-x}$Nb$_x$O$_3$ is marked by a step-like softening, which is significantly smeared out due to the presence of grains in polycrystalline samples. However, the trend that is clearly observed from Fig. 7 is that $T_S$ increases with increasing Nb concentration and goes beyond our experimental range of 380 K when $x$ reaches 0.2. 

\section{discussion}
\subsection{A. Landau theory}
The elastic behavior of a material without any thermodynamic "anomalies" can be modeled using the Varshni function \cite{Varshni}
\begin{equation}
C_{ij}(T) =C_{ij}^0 - {\frac{s}{e^{t/T}-1}}
\centering
\end{equation}
which predicts a gradual stiffening with decreasing temperature that levels off at low temperatures. The elastic behavior of pure ETO obviously deviates from Varshni-behavior.

Landau theory is known to provide a proper framework for a formal analysis of phase transitions \cite{Rettwald1973}, whereby the free energy can be expressed in terms of strain $e_i$ and order parameter components $Q_i$. Slonczewski \textit{et al.} \cite{STOLandau} performed a Landau analysis of the elastic behavior of SrTiO$_3$ in the 1970s, and here we present calculations using the Landau model for the EuTiO$_3$ system, adapting some of the notations used in \cite{STOLandau}. 
Note that inelastic X-ray scattering has indentified zone boundary R-point q=(0.5 0.5 0.5) acoustic phonon softening at this SPT \cite{Inelastic_X-ray_ETO}. 
Thus the amplitudes of the degenerate zone boundary soft acoustic modes $Q_1$, $Q_2$, $Q_3$ can be naturally employed as the order parameter. 
Considering the crystal symmetry, the free energy $F$ should include an antiferrodistortive energy $F(Q_1,Q_2,Q_3)$, the elastic energy $F(e)$, and a contribution from coupling between the order parameter and strain $F(Q,e)$:

\begin{multline}
F(Q_1,Q_2,Q_3) = A_1 (Q_1^2+Q_2^2+Q_3^2 )+A_2 (Q_1^2+Q_2^2+Q_3^2)^2\\
+A_3 (Q_1^2 Q_2^2+Q_2^2 Q_3^2+Q_3^2 Q_1^2 )
\end{multline} 

\begin{multline}
F(e)=\frac{1}{2} {C_{11}^{c} (e_1^2+e_2^2+e_3^2 )}+C_{12}^{c} (e_{1} e_{2}+e_{2} e_{3}+e_{3} e_{1} )\\
+\frac{1}{2} {C_{44}^{c} (e_{4}^{2}+e_{5}^{2}+e_{6}^{2})}
\end{multline} 

\begin{multline}
F(Q,e)=-B_1(e_1Q_1^2+e_2Q_2^2+e_3Q_3^2)\\
-B_2(e_1(Q_2^2+Q_3^2)+e_2(Q_1^2+Q_3^2)+e_3(Q_1^2+Q_2^2))\\
-B_t(e_4Q_2Q_3+e_5Q_3Q_1+e_6Q_1Q_2)
\end{multline} 
Here $C_{11}^c$, $C_{12}^c$ and $C_{44}^c$ are elastic moduli in the cubic phase, $A_1$, $A_2$, $A_3$ are Landau expansion coefficients, and $B_1$, $B_2$, $B_t$  are coupling constants. 

Assuming that the volume strain does not couple to the order parameter, this gives 
\begin{equation}
B_e \equiv -B_2 = +B_1   
\centering
\label{4}
\end{equation}

Under the equilibrium conditions $\frac{dF}{dQ}=0,\frac{dF}{de}=0$
this gives 

\begin{multline}
e_{i}  = \frac{B_{e} (3Q_{i}^{2}  - Q^{2})}{C_{11}^{c}- C_{44}^{c}} (i=1,2,3; Q^2=Q_1^2+Q_2^2+Q_3^2 )\\
e_{4} = \frac{B_{t} Q_{2} Q_{3}}{C_{44}^{c}}, etc 
\end{multline}

The elastic constants are the second derivatives of the free energy $F$ with respect to strain and may be calculated using the Slonczewski-Thomas formalism \cite{STOLandau}

\begin{equation}
C_{ij} = C_{ij}^c - M ^{-1}{\sum_{k}\frac{\partial^2 F}{\partial e_{i}\partial Q_{k}}}{\omega_{k}}^{-2}\frac{\partial^2 F}{\partial Q_{k}\partial e_{j}}
\centering 
\end{equation}
         
Substituting (2) $\to$ (6) into (7) and assuming $Q =(0,0,Q_s )$, we obtain the elastic moduli in the tetragonal phase:

\begin{align*}
C_{11} &= C_{11}^c-D\\
C_{44} &= C_{44}^c-E\\
C_{12} &= C_{12}^c-D\\
C_{33} &= C_{11}^c-4D\\
C_{13} &= C_{12}^c+2D\\ 
C_{66} &= C_{44}^c\\ 
\end{align*}
where $D=(4B_e^2 Q_s^2)/(M{\omega_{3}}^2 )$, and $E=(B_t^2 Q_s^2)/(M{\omega_{2}}^2 )$. \textit{M} is the mass density of the oxygen atoms participating in each optical vibration mode, and $\omega_{i}$ ($i$=1,2,3) are the natural frequencies of these vibration modes. Thus, as the cubic to tetragonal SPT is approached, a step-like softening is expected in the elastic moduli $C_{11}$, $C_{12}$, $C_{44}$, $C_{33}$, step-like stiffening is expected for $C_{13}$,  and $C_{66}$ is expected to follow the trend of the cubic shear modulus $C_{44}$. The step-like softening is clearly observed for $C_{44}$ (Fig. 5), and even though the large absorption below $T_S$ prevents us from determining the full elastic tensor below the transition, the precipitous drop in $C_{11}$ and $C'$ strongly suggests that these moduli undergo a similar step-like softening. While our Landau analysis predicts such behavior for the longitudinal modulus $C_{11}$, $C' =\frac{1}{2}(C_{11}-C_{12})$ is expected to smoothly continue through the SPT. This is clearly not observed. Interestingly, the same discrepancy is found in STO \cite{ElasticModuliSTO}, and remains unexplained.

\begin{figure}[!b]
\centering
\includegraphics[trim = 0mm 0mm 0mm 0mm, clip, width=0.6\textwidth]{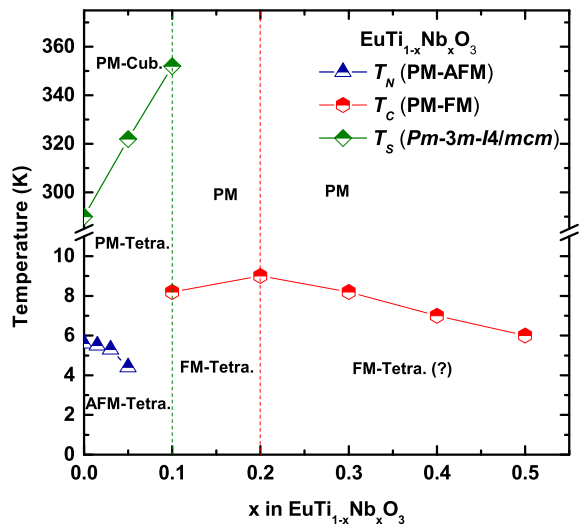}
\caption{The temperature of the structural and magnetic phase transition \textit{vs.} Nb content $x$ in EuTi$_{1-x}$Nb$_x$O$_3$.}
\label{FIG.8}
\end{figure}

\subsection{B. Precursor effect}
In every aspect, the elastic response of ETO through the SPT appears to be very similar to that of STO, which is not surprising in light of the resemblance between their phonon behaviors \cite{STOinelastic,Inelastic_X-ray_ETO}. 
The SPT in both STO and ETO is driven by a condensation of zone-boundary phonon modes, which is indicative of the displacive nature of both SPTs.
From a lattice dynamics point of view, precursor effects have been observed in STO by birefringence measurements \cite{Roleder2012}, where fluctuating ferroelastic clusters form above $T_S$. 
Nevertheless, the TA-TO mode coupling which exists in STO and explains its precursor effects, is actually absent in ETO \cite{Bettis2011}. 
The large neutron absorption inherent to Eu and the difficulty in growing large ETO crystals significantly hamper inelastic neutron experiments and further information regarding the phonon behavior has not yet been obtained. However, our observation of a gentle rounding rather than a sharp step in $C_{ij}$ indicates that the precursor effect takes place in ETO starting about 20 K above $T_S$, where local elastic nanoregions develop. 
This is consistent with the pair distribution analysis \cite{Allieta2012} that showed that the dynamic fluctuations occur above $T_S$, \textit{i.e.} tetragonal twinning segments form along three crystallographic orientations. This unexpected precursor effect in ETO has been analyzed by normalizing the transverse modes and taking the derivatives of the modes with respect to momentum, yielding the softening of the transverse acoustic mode as a function of temperature \cite{ETOprecursor}. The coupling between the TA mode and the elastic constants gives rise to the elastic softening as the SPT is approached. Furthermore, the presence of the precursor effect suggests that the transition is not purely of the displacive type, but includes elements of an order/disorder transition. 

\subsection{C. Effect of Nb doping }
Fig. 8 presents the  EuTi$_{1-x}$Nb$_x$O$_3$ phase diagram with the structural and magnetic transition temperatures as a function of Nb content ($T_N$ for $x$ = 0.015, 0.03 and $T_C$ for $x$ $\ge$ 0.3 samples are taken from Ref. \cite{Li2014}). The magnetic ground state changes from AFM to FM when $x$ $\ge$ 0.1. While $T_N$ gradually decreases from 5.7 K ($x$ = 0) to 4.5 K ($x$ = 0.05), $T_C$ increases slightly from 8.2 K ($x$ = 0.1) to a maximum of 9.3 K ($x$ = 0.2), and decreases with further Nb doping. Structurally, Nb doping enhances $T_C$ from 288 K ($x$ = 0) to 353 K ($x$ = 0.1). Nevertheless, the lattice distortion at 100 K is weakened with increasing Nb content. 
The well established tolerance factor $t$ concept predicts the symmetry of ABO$_3$ perovskites structure, with $t= \frac{r_A+r_O}{\sqrt{2}(r_B+r_O)}$, and $r_A$, $r_B$, and $r_O$ are the ionic radii of each ion. 
The radius of Nb$^{4+}$ ($r$ = 0.68 {\AA}) is greater than that of Ti$^{4+}$  ($r$ = 0.605 {\AA}) \cite{Shannon1976}, which means that substituting Nb for Ti reduces the tolerance factor and promotes the oxygen rotating instabilities. 
This is in favor of the cubic to tetragonal transformation occurring at higher temperatures with increasing Nb doing, as confirmed by RUS results. 
Nevertheless, this tolerance factor model, exclusively based on ion sizes, becomes problematic in an attempt to explain the difference in $T_S$ between STO and ETO. Neither does it explain the weaker octahedral distortion in EuTi$_{1-x}$Nb$_{x}$O$_{3}$ with increasing Nb concentration. 
It has been suggested that the high $T_S$ of ETO compared to STO is associated with the hybridization between Eu $f$ states and Ti $d$ orbitals, which is also believed to be the origin of the giant spin-lattice coupling present in ETO \cite{Akamatsu2012}. 
Previous studies have shown that Nb doping enhances $T_S$ of STO to higher temperatures due to the local changes in the hybridization between the (Ti, Nb)-$d$ orbitals and $O$-2$p$ orbitals \cite{STO1978}. Another important factor is the covalency of the transition metal-oxygen bond \cite{GarciaFernandez2010, Cammarata2014}. Ab initio calculations have shown that the octahedral tilting angle in KMF$_3$ (M = transition metal) decreases linearly with the electron occupancy of the $\pi$-bonding $t_{2g}$ orbitals of the transition metal. In the present system, Ti$^{4+}$  has a 3$d^0$ configuration. In contrast,  Nb$^{4+}$ possesses one electron in the $t_{2g}$ orbitals, which most likely alters the covalency of the (Ti,Nb)-$O$ bond and results in a weaker lattice distortion at low temperatures. It is noteworthy that EuNbO$_3$ is reported to remain cubic at room temperature, and octahedral tilting does not occur \cite{EuNbO3}.

\section{summary}
In conclusion, we have investigated the structural and magnetic transitions in both polycrystalline and single crystal samples of EuTi$_{1-x}$Nb$_x$O$_3$ (0 $\le x \le$ 0.3) using synchrotron powder XRD, resonant ultrasound spectroscopy, and magnetization measurements. 
The cubic ($Pm\bar{3}m$) to tetragonal ($I4/mcm$) structural phase transition in pure and doped EuTiO$_3$ is characterized by a step-like elastic softening in $C_{44}$, $C_{11}$, and $C' = \frac{1}{2}(C_{11}-C_{12})$. 
Nb substitution for Ti significantly enhances $T_S$ to higher temperatures ($T_S$ = 322 K and 353 K for EuTi$_{0.95}$Nb$_{0.05}$O$_3$ and EuTi$_{0.9}$Nb$_{0.1}$O$_3$, respectively). 
In contrast, the octahedral tilting angle decreases with increasing Nb doping, which is possibly caused by the introduction of $t_{2g}$ electrons on the Ti site. 
In the meanwhile, the introduction of itinerant $t_{2g}$ electrons also switches the magnetic ground state of EuTiO$_3$ from AFM to FM. 
Our study emphasizes the important role of the covalency of the transition metal - oxygen bond in determining the octahedral tilting amplitude and magnetic ground state in the present system. 
Band structure calculations of EuTi$_{1-x}$Nb$_{x}$O$_{3}$ will be useful to further examine how the hybridization between the Eu-$f$, (Ti,Nb)-$d$ and O-$p$ orbitals changes with electron doping concentration on the transition metal site.

\section{acknowledgment}
L.L. acknowledges useful discussions with Dr. A. Bussmann-Holder. This research (L.L. and D.G.M.) is funded in part by the Gordon and Betty Moore Foundation’s EPiQS Initiative through Grant GBMF4416. 
J.R.M. and J.-Q. Yan acknowledge support from the U.S. Department of Energy, Basic Energy Sciences, Materials Sciences and Engineering Division.
Z.L.D and H.D.Z are supported by NSF-DMR-1350002. 
Use of the Advanced Photon Source at Argonne National Laboratory was supported by the U. S. Department of Energy, Office of Science, Office of Basic Energy Sciences, under Contract No. DE-AC02-06CH11357.

\bibliography{ll}

\end{document}